\documentstyle[referee]{mn}

\input{psfig.sty}
\newcommand{\be}{\begin{equation}}    
\newcommand{\ee}{\end{equation}}
\newcommand{\beq}{\begin{eqnarray}}
\newcommand{\eeq}{\end{eqnarray}}
\newcommand{\beqn}{\begin{eqnarray*}}
\newcommand{\eeqn}{\end{eqnarray*}}

\def\lappreq{\! \stackrel{\scriptscriptstyle <}{\scriptscriptstyle
\sim}\!}
\def\gappreq{\!\stackrel{\scriptscriptstyle >}{\scriptscriptstyle \sim}\!}

\begin{document}



\title[GWs from newly born, hot NSs]
{Gravitational waves from newly born, hot neutron stars}
\author[V. Ferrari, G. Miniutti, and J.A. Pons]
{V. Ferrari$^1$, G. Miniutti$^1$ and J.A. Pons$^{1,2}$ \\ $^1$ Dipartimento
di Fisica ``G.Marconi", Universit\` a di Roma ``La Sapienza" and Sezione
INFN  ROMA1, \\ Piazzale Aldo Moro 2, I-00185 Roma, Italy \\ $^2$
Departament d'Astronomia i Astrof\'{\i}sica, Universitat de Val\`encia,
46100 Burjassot, Val\`encia, Spain
}

\maketitle

\begin{abstract}
{We study the  gravitational radiation associated to 
the non--radial oscillations of newly born, hot neutron stars.
The frequencies and damping times of the relevant
quasi--normal modes are computed for two different models of 
proto--neutron stars, at  different times of evolution,
from its birth until it settles down as a  cold neutron star.
We find that the oscillation properties of proto--neutron stars are 
remarkably different from those of their cold, old descendants  and that 
this affects the characteristic  features
of the gravitational signal emitted during the post-collapse evolution. 
The consequences on the observability of
these signals by resonant--mass and interferometric
detectors are analyzed.  We find that gravitational waves from the 
pulsations of a newborn 
proto--neutron star in the galaxy could be detected
with a signal to noise ratio of 5 by the first generation
interferometers, if the energy stored in the modes
is greater than  $\sim 10^{-8}~M_\odot c^2$,
or by a resonant antenna if it is greater than $\sim 10^{-4}~M_\odot c^2$.
In addition since at early times the frequency of the spacetime modes is
much lower than that of a cold neutron star, they would be also detectable 
with the same signal to noise ratio if a comparable amount of energy 
is radiated into these modes. 
}

\end{abstract}



\section{Introduction}

According to the standard theory of star formation, neutron stars are born in the aftermath of 
successful core--collapse supernova explosions, as the stellar remnant becomes 
gravitationally decoupled from the expanding ejecta. 
Following the gravitational collapse, the proto--neutron star (PNS) radiates
its binding energy (about 0.1 $M_\odot$)  via neutrino emission 
in a timescale of tens of seconds.
A small fraction of this energy reservoir is radiated in gravitational waves (GWs),
and it is interesting to establish for how long, and at which
frequencies  the gravitational radiation will be emitted. 
The best currently available  2D-numerical simulations of a core collapse,
and the most reliable estimates of gravitational wave emission 
consider, at best,  the first hundred milliseconds from the
collapse onset. In a number of axisymmetric simulations
\cite{zwerger,harry1,freyer}, 
the flux of gravitational waves is evaluated through the quadrupole formalism,
by using the time variation of the energy density of the collapsing system obtained 
either with Newtonian or  general relativistic hydrodynamical codes.
The emitted energy is estimated to range between $10^{-8}$ and $10^{-6} M_\odot$.
Full three dimensional simulations in general relativity are not yet 
available, but there are some Newtonian studies \cite{rampp,brown} that, by assuming
small deviations with respect to axial symmetry,  
show similar results. 
In any event, the  compact object which forms after the collapse possesses a certain
amount of mechanical energy, which will partially be released  through violent 
oscillations in the form of gravitational radiation. 

The spectrum of the non--radial oscillations of a neutron star (NS) has a
very rich structure that depends on how the temperature and the internal
composition vary along its history. For instance, the frequency of
the fundamental  mode  has been found to scale approximately 
as $\sim \sqrt{M/R^3}$, where $M$ is the gravitational mass
of the star and $R$ is its radius, whereas that of the acoustic
p--modes scales linearly with $M/R$. Since 
during the first tens of seconds of its life the star contracts
and its gravitational mass decreases, 
one expects significant changes in the frequency of these modes.
In addition,  new families of modes may appear or disappear for several
different reasons: {\it i)} as the star's entropy gradient decreases with time,
the frequency of the gravity modes (g--modes) that are present at finite
temperature migrate toward zero; {\it ii)}  if no significant diffusion mechanism is present,
layers of different chemical composition accumulating on the stellar surface
would produce one or more density discontinuities, which  would
give rise to  low frequency (typically $< 200$ Hz) discontinuity g--modes;
{\it iii)} if the density of the inner core becomes high enough, a phase transition
may occur, involving pion and/or kaon condensation as well as 
a transition from ordinary nuclear matter to quark matter \cite{Pra97}
which may produce a density discontinuity in the inner core of the star. In this case, a new
discontinuity g--mode would appear, at a frequency higher than  the
previous ones - typically  $ > 700$ Hz - \cite{sotani,tutti4}; 
{\it iv)}  in previous works
\cite{reise92,lai94} it has been shown that g-modes due to composition gradients
in the inner core of cold NSs have typical frequency below $200~$Hz.
In hot PNS, the composition gradients 
are larger and produce g-modes of higher frequencies, even at constant entropy.
{\it v)} Finally, it should be mentioned that relativistic stars do also possess 
a family of modes,  called w--modes,
that  are associated to perturbations of the gravitational field,
and couple very little (or do not couple at all, as for the axial w--modes) 
with the motion of the fluid \cite{chferax,ks92}.  
The properties of the w--modes depend on the internal composition of the
star \cite{nilsprl,omar}, and therefore they also change as the star evolves.

In this paper we study how the characteristic frequencies and damping times of 
some relevant classes of quasi--normal modes (QNMs) of a newly born, 
hot PNS change during its early evolution until it settles as a cold, catalyzed NS. 
Despite the fact that PNSs are expected to be born with a significant amount of angular
momentum, in this work we neglect the effects of rotation. In this way, we
can isolate and better understand the effects of the thermal and chemical
evolution, and we can also clearly indicate the differences between hot
PNS and cold NS oscillation properties. This study gives an interesting
indication of the frequency range where the gravitational signals emitted
immediately after gravitational collapse should be looked for. The
paper is organized as follows. In Section 2 we review the evolutionary
stages immediately following the birth of a proto--neutron star and discuss the
dissipative precesses and timescales relevant during the Kelvin-Helmholtz
epoch. In Section 3, we remind the physical processes that produce the gravity
modes, and shortly describe the procedure we use  to compute the QNM frequencies.
The results of our study are presented in Section 4, 
where we also  discuss  the detectability
of the gravitational signal associated to the quasi--normal modes by
resonant bars or ground based interferometric detectors.
The main findings of the paper are summarized in the concluding remarks.

\section{The first minute of a neutron star life.}

\begin{figure}
\begin{center}
\psfig{figure=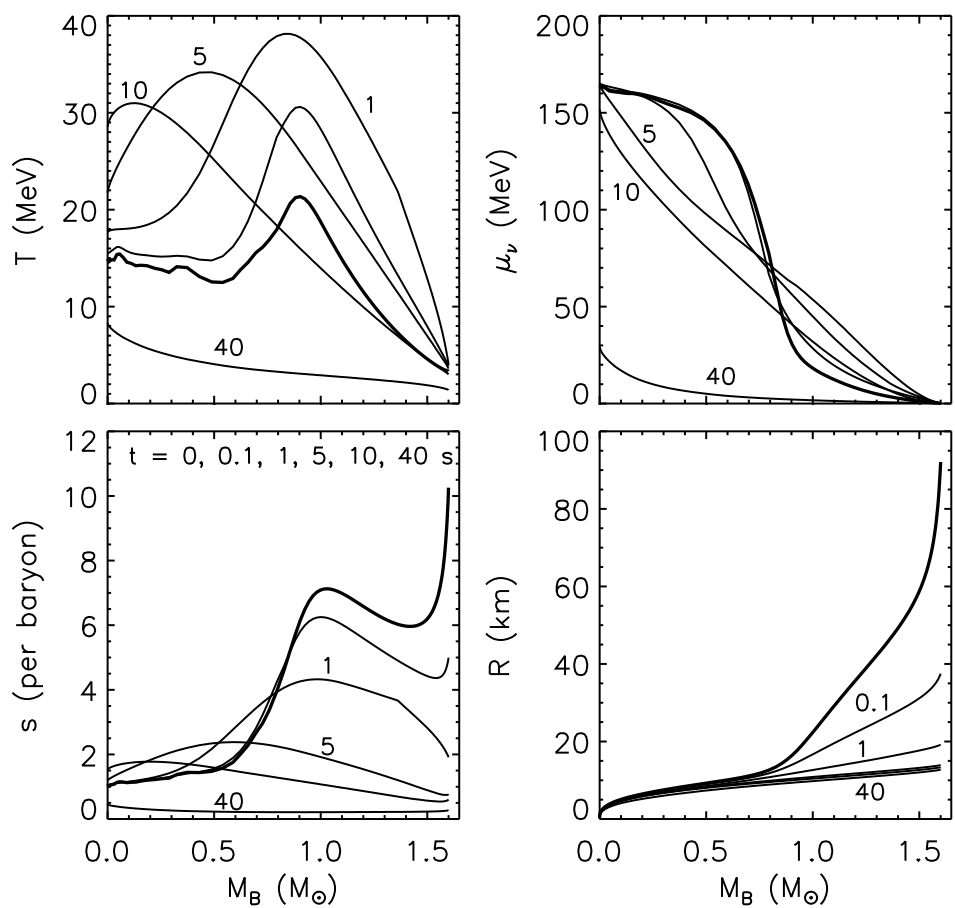,width=14cm}
\vspace{0.5cm}
\caption{Temporal evolution of the temperature $T$, entropy per baryon $s$, 
neutrino chemical potential $\mu_\nu$, and radius $R$, as a function of
the enclosed baryonic masses $M_B$ for model A. The thick line is the
initial profile and the figure displays a temporal sequence corresponding
to $t= 0.1, 1, 5, 10$ and 40 seconds of evolution.} 
\label{FIG1}
\end{center}
\end{figure}

A newly born PNS is very different from its
cold descendant, i.e. a NS. Indeed, initially, the PNS is hot,
lepton rich, and optically thick to neutrinos, which are temporarily
trapped within the star. The subsequent evolution  is dominated
by neutrino diffusion which first deleptonizes and
later cools the star \cite{bl86}. The evolution of a PNS proceeds
through several distinct stages and has been recently studied in great
detail, including the possibility of developing a core of exotic matter
such as hyperons, a kaon condensate,  or  quark matter.
The results we present in this paper are based on simulations of PNS
evolution  described in Pons et al. (1999; 2000; 2001). 
We consider two models of PNS evolution, that use
two different  equations of state (EOSs). 
In both cases the EOS of baryonic matter is a finite-temperature, field-theoretical model
solved at the mean field level, where the interactions between
baryons are mediated by the exchange of $\sigma$, $\omega$ and $\rho$ mesons. Electrons and
muons are included in the model as non interacting particles, being the contribution due to
their interactions much smaller than that of the free Fermi gas. 
We consider two of the models of evolving PNS studied by Pons et al. (1999; 2001)
where neutrino transport is treated using  the
diffusion approximation, that allows to obtain global properties of the neutrino
emission, like average energies, integrated fluxes and timescales.
Model A has a soft EOS \cite{GM91} while
model B is stiffer \cite{MS96} and allows for the formation of a quark core at some point of 
the evolution. The mixed phase of baryons and quarks is
constructed to satisfy Gibbs' phase rules for mechanical, chemical and thermal
equilibrium, assuming global charge neutrality.  
The quark matter EOS is computed using the MIT bag model (with Bag constant B= 150
MeV/fm$^3$). 
A detailed discussion on the EOS, on the evolution of
PNSs and on the  microphysical inputs (opacities) can be found 
in the above references. 
However, here it is useful to summarize the different stages of the early evolution
of a PNS referring, as an illustrative example, to Model A.
In Figure \ref{FIG1} we show some relevant physical variables that
characterize the PNS  status  at different instants of  time ($0$,
$0.1$, $1$, $5$, $10$, $40$ seconds): the entropy per baryon, $s$, the
temperature $T$, the electron neutrino chemical potential  $\mu_\nu$, and
the radius $R$. Each quantity is plotted versus the enclosed baryonic
mass, $M_B$.

The different stages of the early evolution can be summarized as
follows.

i) Immediately after the core bounce and the passage of the shock wave
through the outer PNS mantle, the star contains an unshocked, low
entropy core of mass $\simeq 0.7$ M$_\odot$ in which neutrinos are
trapped. The core is surrounded by a low density, high entropy 
mantle (see Figure 1, left--bottom)
that is accreting matter falling through the shock from the outer iron shell 
and it is rapidly losing energy due to electron captures
and thermal neutrino emission. The mantle extends up to the shock, which
is temporarily stalled at a radius of about $200$~km prior to an eventual
explosion.  If the supernova is successful, after a few hundred milliseconds 
accretion becomes unimportant  and the shock lifts off the
stellar envelope. In a few tens of seconds, extensive neutrino losses reduce the lepton
pressure provoking the contraction of the mantle. The radius of the PNS is now
about $20-30$~km.

ii) Neutrino diffusion deleptonizes the core on a time scale of
$10-15$~s. The diffusion of degenerate, high-energy ($200$~MeV) neutrinos
from the core to the surface, from where they escape as low-energy neutrinos
($10-20$~MeV), generates a large amount of heat within the star,
resulting in temperatures up to $50$~MeV, 
while the core entropy approximately doubles (see Figure 1, left panels).
The inverse temperature gradient which arises in the inner
region produces a heat wave that propagates inward. At the end of the
deleptonization epoch, the star reaches its maximum central temperature.

iii) After  approximately $15$ s,
the PNS has become  lepton-poor but it is still hot.  
The net number of neutrinos in its interior is low, but
thermally produced neutrino pairs of all flavors are abundant and dominate
the emission. Thus,  neutrino diffusion continues to cool the star, while the
average neutrino energy decreases, and the neutrino mean free path
increases.  After approximately $50$ seconds, the mean free path becomes
comparable to the stellar radius, and the star finally becomes transparent
to neutrinos. By this time, the temperature has dropped to $1-5$~MeV
($\approx 10^{10}$ K) and the star has radiated almost all of its binding energy.
A neutron star is born.
\subsection{Dissipation time scales}
Since the focus of this paper is on the emission of gravitational
radiation produced by the  violent pulsations of a newly born PNS,
it is crucial to understand whether these
oscillations may be  damped by other dissipative processes acting on
timescales shorter than the typical  gravitational  timescales.

We have seen that in a PNS the energy  is
transferred and redistributed dominantly by neutrino diffusion. Hence, when
one considers hydrodynamical processes, the relevant dissipation
time-scales are those corresponding to the neutrino viscosity,
diffusivity, thermal conductivity, or thermodiffusion, rather than those associated to
the matter kinetic coefficients. Transport properties of degenerate neutrinos
in dense matter were studied in detail by Goodwin \& Pethick (1982) and by
van den Horn \& van Weert (1984). They found that for the thermodynamical
conditions typical in PNSs and for  strongly degenerate neutrinos ($\mu_\nu \gg k_B T$)
the dominant dissipative process is neutrino chemical diffusion, followed
by thermal conduction and viscosity. Thus, the hierarchy of dissipation
timescales is $\tau_D < \tau_C < \tau_{visc},~ $ where
 $\tau_D$,  $\tau_C$, and  $\tau_{visc}$ are the diffusion, thermal
conduction, and viscous timescales, respectively. 
$\tau_D$ can be estimated by the following approximate relations \cite{vv84}
\be
\tau_D({\rm s}) \approx 24~s~n_B^{1/3} \left(\frac{0.01}{Y_\nu} \right)
\left( \frac{T}{10 {\rm MeV}} \right)
\left( \frac{\mu_\nu}{100 {\rm MeV}} \right),
\label{taud}
\ee
where  $s$ is the entropy per baryon in units of the Boltzmann
constant, $n_B$ is the baryon density
(in fm$^{-3}$), and $Y_{\nu}$ is the neutrino fraction. 

Conversely, if neutrinos
are non-degenerate ($\mu_\nu \ll k_B T$), the hierarchy between neutrino
diffusion and thermal conduction is reversed and $\tau_C < \tau_D < \tau_{visc} $.
In this case  $\tau_C$ can be expressed as
\cite{vv84}
\be
\tau_C({\rm s}) \approx 1.5~s~n_B^{1/3}  \left(\frac{0.01}{Y_\nu} \right)
\left( \frac{T}{10 {\rm MeV}} \right)^2,
\label{tauc}
\ee
In both cases, the timescale
$\tau_{visc}$ is a factor $10-50$ longer.

\begin{figure}
\begin{center}
\psfig{figure=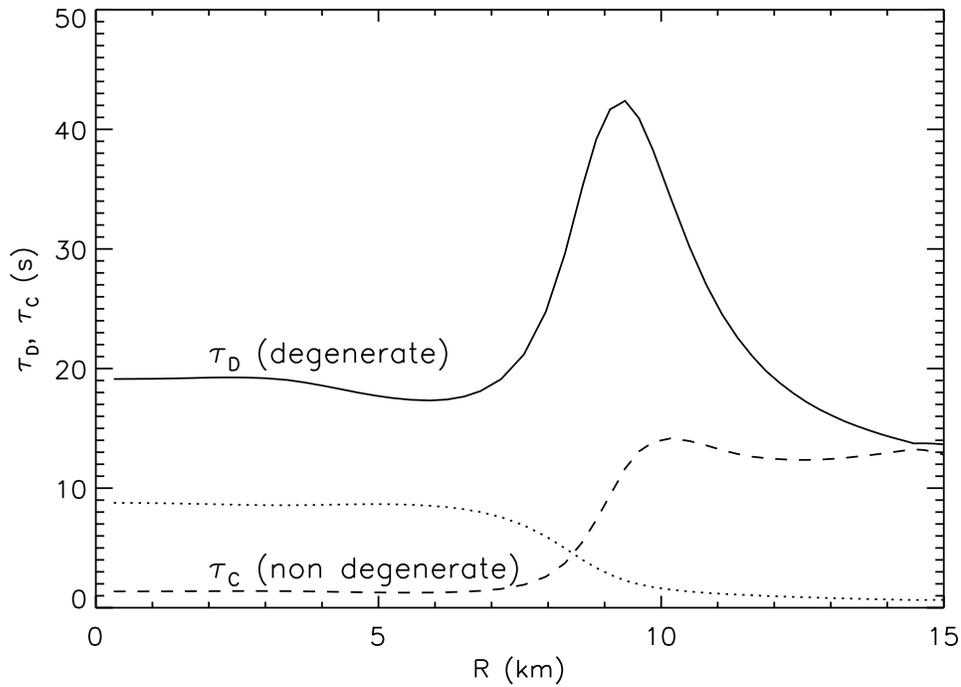,width=14cm}
\vspace{0.5cm}
\caption{Neutrino dissipation timescales corresponding to diffusion ($\tau_D$)
and thermal conductivity ($\tau_C$) at $t=0.25$ s for model A in the inner
$15~$km of the PNS. The diffusion timescale (solid line) dominates in the
degenerate regime, while the conductive timescale (dashes) applies to
matter with non--degenerate neutrinos. For reference, we also show the
dimensionless neutrino degeneracy parameter, $\eta=\mu_\nu/T$ (dotted
line), which indicates that the regime is degenerate from the star center
up to about $10$ km, while it is non degenerate in the outer, low density,
region.
}
\label{figtau}
\end{center}
\end{figure}

In summary, during the first $10-20$ seconds of the star life,
diffusion is the dominant dissipative process in the inner regions
(where neutrinos are degenerate),
while thermal conduction dominates in the outer region. 
Afterwards, thermal conduction becomes the
more efficient mechanism of energy dissipation until the matter composing
the star becomes transparent to neutrinos.
The order of magnitude of both
$\tau_D$ and $\tau_C$ varies from a few seconds to tens of seconds, and it is
comparable to the overall evolutionary timescale. This is shown in Figure
\ref{figtau}, where the diffusion (solid line) and  the conductive (dashed
line) timescales are plotted  as a function of the radius for
model A, and $t=0.25$ seconds after the birth of the PNS. The dotted line is the 
degeneracy parameter of
neutrinos ($\eta = \mu_\nu / T$), and it is shown to indicate in which
region  $\tau_D$ and $\tau_C$ dominate. As a rule of
thumb, $\tau_D$ must be used when $\eta>1,$ while $\tau_C$ is the relevant
timescale when $\eta<1$. 
In conclusion, a conservative estimate of the dissipative timescales
before neutrino transparency is
$\tau_D  \approx \tau_C \approx 10-20$ seconds.
In the following we will denote by $\tau_{diss}$ the minimum of the typical
dissipative timescales,  $\tau_{diss}={\rm min}( \tau_{D}, \tau_{C}, \tau_{visc})$.

\section{Quasi-normal modes of relativistic stars}

In order to  understand what are the differences between the oscillation properties of
PNSs and those of old, cold NSs, we shall briefly recollect some results
obtained in the literature for  cold NSs.  
A systematic analysis of the pulsation properties of cold NSs has been done by Andersson and
Kokkotas (1998) for a large range of EOSs; they found  a set of empirical 
relations that connect the mode frequencies
and damping times to the mass and radius of the star.
As an example, from their relations one finds that
for a cold NS with $~M=1.4 M_\odot~$ and $~R = 10~$ km  the  inferred values of the 
frequency and damping times of the f--mode, and of the first p-- and w-- modes 
for $l=2$ are
\begin{eqnarray}
\nu_f  &\approx& 2~{\mathrm kHz} \ ,
\qquad\quad \nu_{p_1} \approx 7~{\mathrm kHz}  \ ,
\qquad \nu_{w_1} \approx 11.8~{\mathrm kHz} \ ,\nonumber \\
\tau_f &\approx& 10^{-1}~{\mathrm s}  \ ,
\qquad\quad   \tau_{p_1} \approx  5~{\mathrm s}  \ ,
\qquad \;\;\;\; \tau_{w_1} \approx 2.4\times 10^{-5}~{\mathrm s}\ .
\label{nilsfits}
\end{eqnarray}
A further family of modes which may appear in NSs are
the so-called gravity or g--modes, whose restoring force is buoyancy.  
It is known \cite{TH66} that the radial
acceleration of a fluid element displaced by its equilibrium position can be written
as
\begin{equation}
a = - \frac{e^{-\lambda/2}}{(\epsilon +p)\gamma p}\,\left(-\frac{dp}{dr}\right)\,
S(r)\,\Delta r \ ,
\label{accrad}
\end{equation}
where $\epsilon$ and $p$ are the energy density and the pressure, $\Delta r$ is
the radial displacement of the fluid element, 
$\gamma = [(\epsilon+p)/p ] [\partial p/ \partial \epsilon ]_{s,Y_L}$
is the adiabatic index, with $Y_L$ being the lepton fraction, 
$-e^\lambda$ is the $g_{rr}$ component of the unperturbed  metric tensor,
and 
\begin{equation}
S(r) = \frac{dp}{dr} - \frac{\gamma p}{\epsilon +p}\,
\frac{d\epsilon}{dr}
\label{Sdiscr}
\end{equation}
is the relativistic Schwarzschild discriminant.
If $S(r)>0$, the displaced fluid element feels a force that tends to 
restore the initial equilibrium position and the star (or the 
region inside the star) is stable against convection. In this case, a
non--zero frequency spectrum of stable g--modes  appear.
Conversely, if $S(r) <0$, the force tends to move the fluid element away from equilibrium
and the star becomes unstable against convection. 
Finally, if $S(r) = 0$, all g--modes are degenerate at zero 
frequency.

By introducing the  adiabatic sound velocity, $c^2_s$,
which is defined at constant entropy $s$ and chemical composition, and
the so--called equilibrium velocity, $c^2_0$,
\begin{equation}
c^2_s = \left(\,\frac{\partial
    p}{\partial\epsilon}\,\right)_{\rm{s,Y_L}} 
=\frac{p}{\epsilon +p}\,\gamma \ , \qquad
c^2_0 = \frac{dp/dr}{d\epsilon /dr} \ ,
\label{cs}
\end{equation}
the Schwarzschild discriminant (\ref{Sdiscr}) can be cast into  a
form which will be useful   in following discussions
\begin{equation}
\label{discrim}
S(r) = \frac{dp}{dr}\,\left(\,1 - \frac{c^2_s}{c^2_0}\,\right) \ .
\end{equation}
Thus, a stable  g--mode arises only if $c^2_s > c^2_0$, 
and the buoyancy which provides the restoring force can be induced either by entropy
and/or composition gradients \cite{mcd88,reise92,lai94}, by density
discontinuities in the outer envelopes of NSs \cite{finn87,stro93},
or  by density  discontinuities
in the inner core as a consequence of phase transitions at high density
\cite{sotani,tutti4}. The frequency of the g--modes is  always lower than that of the f--mode, 
it is roughly proportional to the Schwarzschild discriminant,
and depends on the physical process  which is source of the buoyancy.
In general, the g--modes of  cold NSs have a typical frequency of less than 200-300 Hz
and very long damping times (typically, $~\tau_g > 10^6~$ s).
We shall see that the situation is  different during the first second of life of a PNS.

\subsection{Computation of the modes}

In this subsection we clarify some technical aspects of the computational
procedure we use to find the frequencies and damping times of the 
quasi--normal modes.  We integrate the equations
describing the polar, non radial perturbations of a non rotating star in
general relativity. We use the same formulation of the perturbed equations 
as in Miniutti et al. (2002), based on  the work of
Detweiler \& Lindblom (1985), who
derived a fourth order system of equations for the metric and fluid
perturbations inside the star. However, we do not use their original
method, because we integrate the system in the  complex frequency domain directly,
rather than working in the  real frequency domain. The system of perturbed equations
is closed by the equation of state. In particular, the equilibrium profiles 
of the energy density and of the sound velocity  are sufficient to
completely solve the perturbation equations in the stellar interior. The
profiles we use are the same as those obtained in Pons et al. (1999; 2001).

The set of equations, that couple the
perturbations of the fluid with those of the gravitational field, are
Fourier expanded and separated by expanding the perturbed tensors in
tensorial spherical harmonics. For each assigned value of the harmonic index
$~l~$ and of the (complex) frequency $~\nu~$, the system of equations
admits only two linearly independent solutions regular at $~r=0$. The
general solution can be found as a linear combination of the two, such
that the  Lagrangian perturbation of the  pressure, $~\Delta p~$, vanishes
at the surface.
Outside the star, the fluid perturbations vanish and the system reduces to
the Zerilli equation \cite{zerilli}. A quasi normal mode of the star is
defined to be a solution of the perturbed equations belonging to a complex
eigenfrequency, which is regular at the center,
continuous at the surface, and which behaves as a pure outgoing wave at
infinity. To compute the QNM complex eigenfrequencies we follow the
method developed by Leins, Nollert and Soffel (1993) who derived a
procedure (the continued fractions method) to evaluate a  function
which is proportional to the ingoing part of the wave, and therefore
vanishes when $~\nu~$ is an eigenfrequency of a QNM.
A very clear account of the method can be found in Sotani, Tominaga \& Maeda (2002).

\section{The gravitational wave spectrum of a proto-neutron star.}

Using the method described in Section 3, we have computed the 
quasi--normal mode frequencies and damping times of the two models of
proto--neutron star, at different instant of time
from $~t=0.2~$s to $~t=50~$s after their formation.
The configurations we consider as initial refers to
$~t=0.2~$s for both models, because at earlier times the PNS evolution
could be affected by dynamical processes like accretion, and
cannot be accurately approximated by a quasi--stationary series of
solutions of the hydrostatic  equations. Thus
we exclude from our study the collapse, the bounce and the first two hundred
milliseconds after formation, for which more detailed hydrodynamical
simulations are needed. 

At $~t=0.2~$s, model A has
$~M=1.58~M_\odot~$ and $~R=34.2~$km, while model B has  the same mass
and $~R=35.1~$km.  Both
models are chosen such that the final stage of the PNS evolution, at
$~t=50~$s in our simulations, represents a ``canonical'' NS with a mass of
about $~M=1.46 M_\odot~$ and a radius of  $~R  \sim 12-13~$ km.
The gravitational mass difference between the initial and final configurations 
($~\approx 0.12 M_\odot~$)
has been  taken away by neutrinos during the PNS evolution.
In the following, 
we shall discuss the behavior of  the frequencies of the fundamental mode,
and the first g--, p-- and w--modes for $l=2$, as the star evolves and cools down.
To identify the different classes of fluid  modes we study the behavior
of the Lagrangian radial displacement.
The eigenfunction of the f--mode has no radial nodes, while the 
g$_1$-- and p$_1$--modes have one radial node.
The w$_1$--mode is easily identified, because it produces  negligible motion in the fluid,
and because its damping time is much shorter
than that of the fluid modes. 

\begin{figure}
\begin{center}
\centerline{\mbox{ 
\psfig{figure=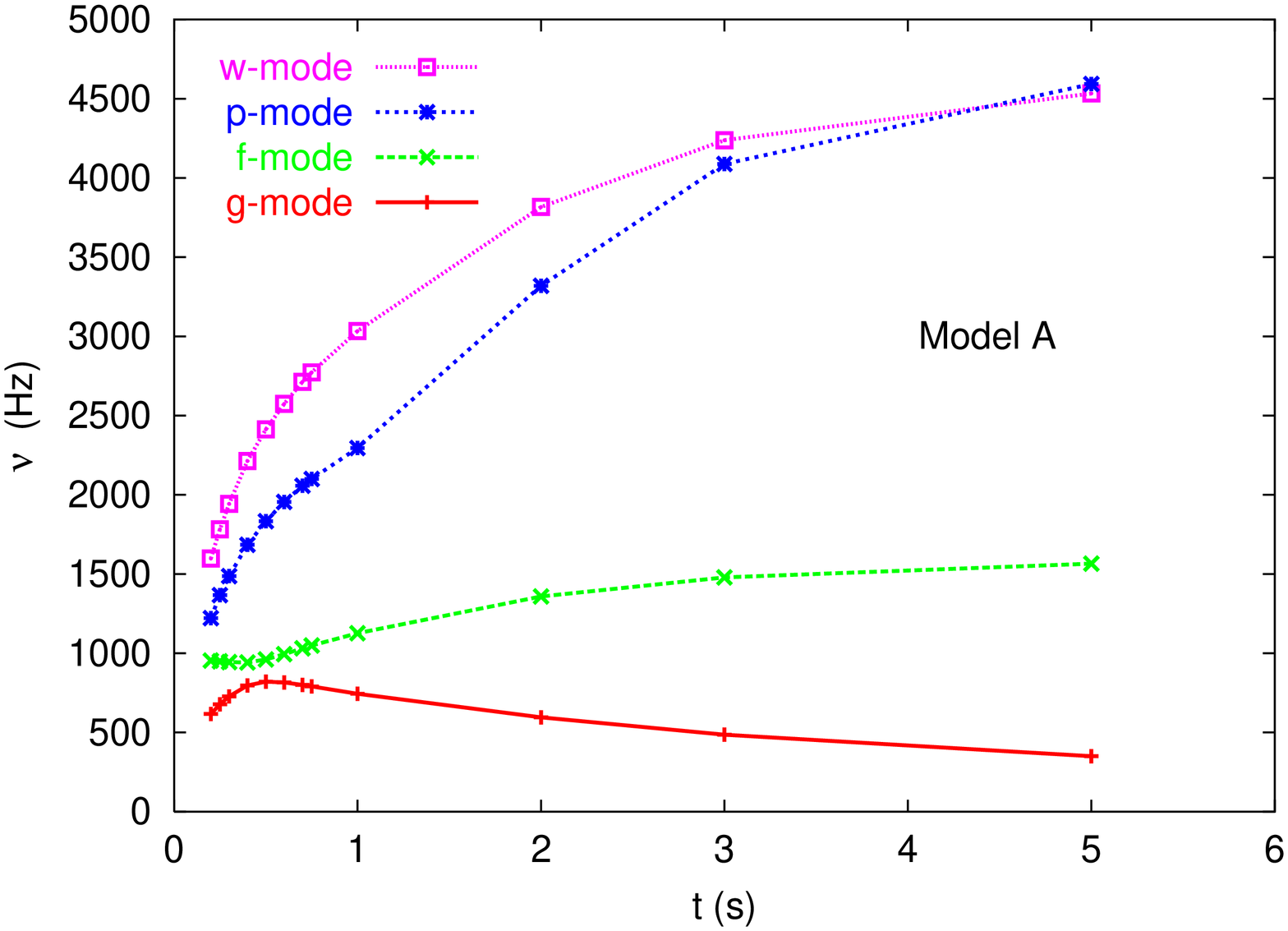,width=9.0cm,height=8.0cm,angle=-90}
\psfig{figure=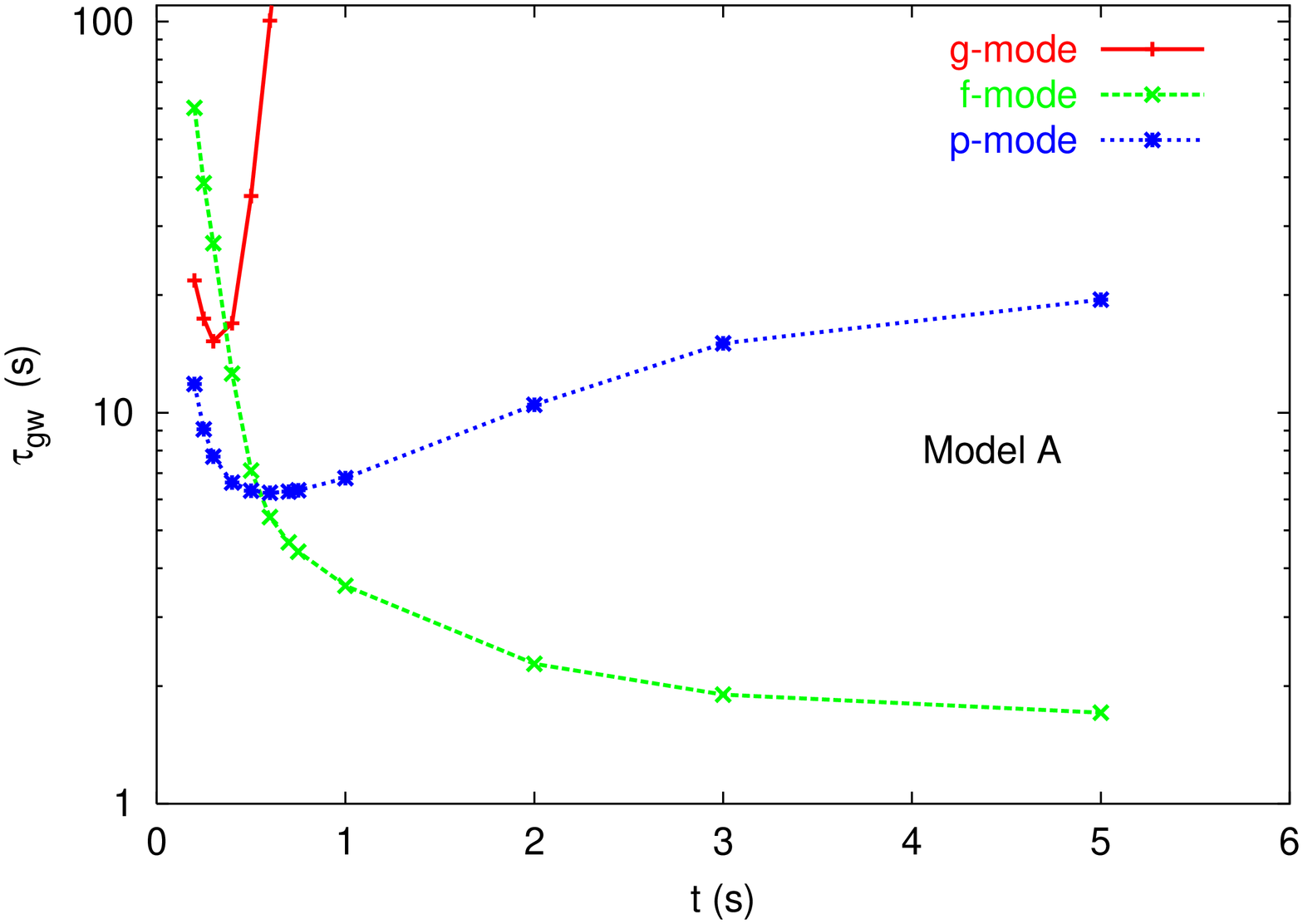,width=9.0cm,height=8.0cm,angle=-90} 
}}
\centerline{\mbox{
\psfig{figure=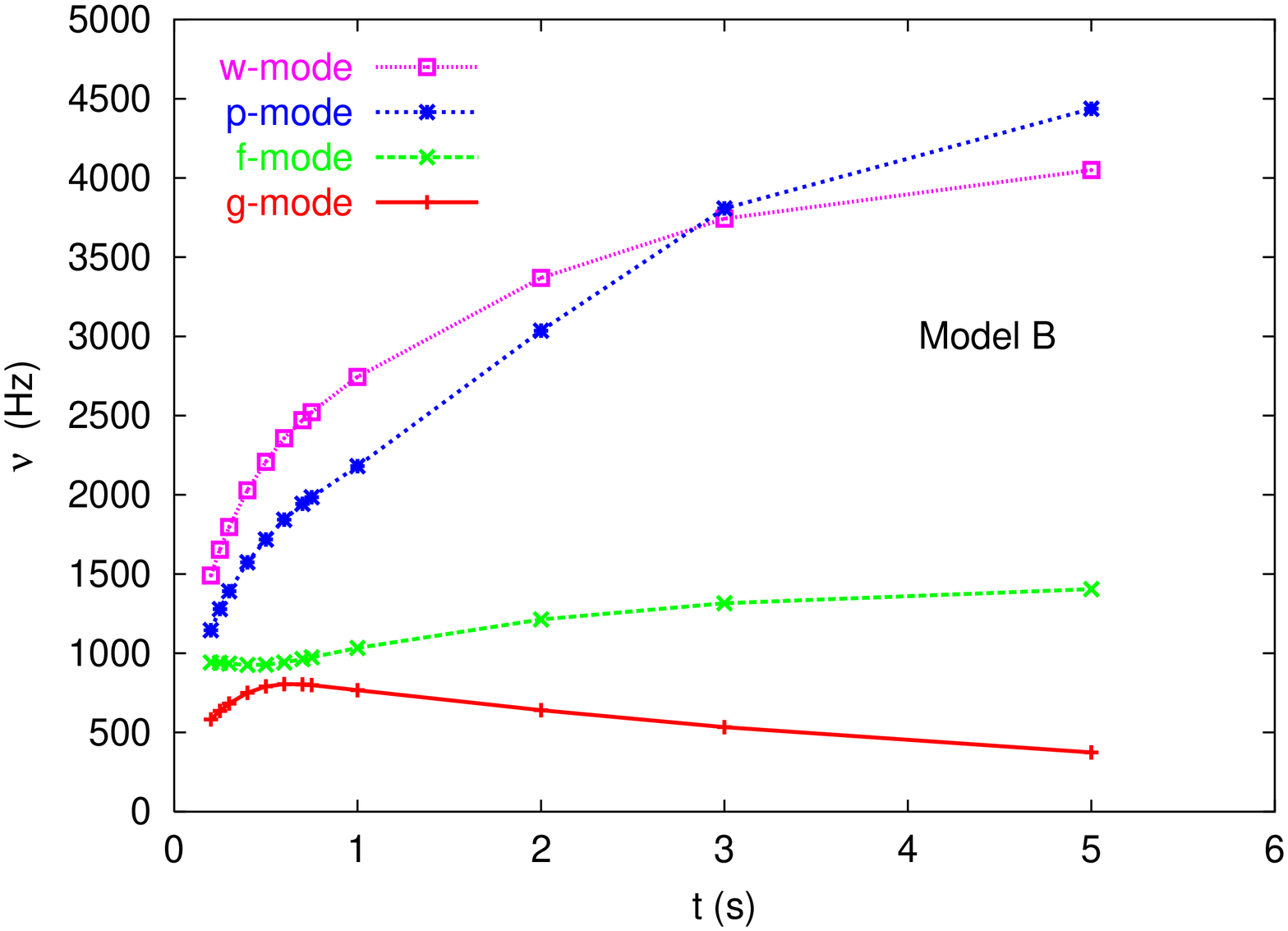,width=9.0cm,height=8.0cm,angle=-90}
\psfig{figure=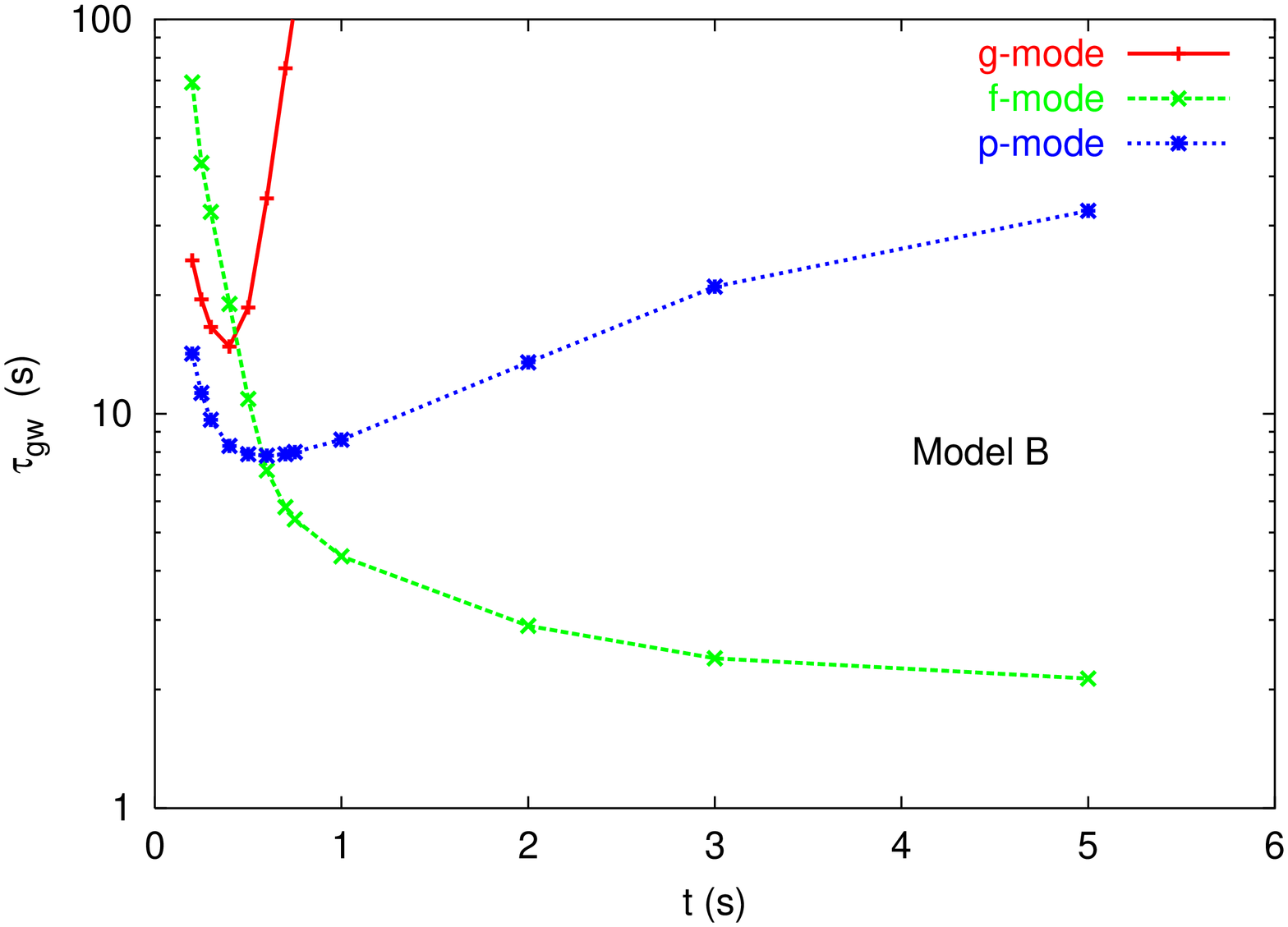,width=9.0cm,height=8.0cm,angle=-90}
}}
\vspace{0.5cm}
\caption{The frequencies and damping times of the lowest QNMs of the proto--neutron star are
plotted as a function of the time elapsed from the gravitational collapse,
during the first $5$ seconds of evolution.
The data plotted in the upper panel refer to model A 
(see Table \ref{table1}) while the lower panel refers to model B
(see Table \ref{table2}), whose main difference from model A is the appearance
of a quark core at some instant of the evolution.
Note that the overall behavior is the same for both models,
the frequency of all the modes clusters in a very narrow range at the
beginning of the PNS evolution and begins to differentiate
after less than $1$ second. After a few seconds the behavior is
already well established and the QNM frequencies tend to their values
for old, cold NSs.
Note also  that the g--mode damping time reaches a deep minimum at about
$t=0.3$ s.  The gravitational damping time of the w--mode is not
shown because it is too small with respect to the scale of the figure,
$~\tau_w = [1.4 - 2.1]\cdot 10^{-4}~$ s.
} 
\label{FIG3}
\end{center}
\end{figure}

The QNM frequencies and damping times for model A and B are tabulated in
Tables \ref{table1} and \ref{table2} at different values of the evolution
time.  The oscillation properties
within the first $~5~$ seconds of the PNS evolution are summarized in
Figure \ref{FIG3}  for  both models.

We find that during the first seconds of life of the PNS the 
frequencies of the f--, p$_1$-- and w$_1$-- modes
are much lower than those of the  cold neutron star which results at the
end of the evolution.  Initially, they cluster in a narrow range within $[900-1500]$ Hz,  
and if we compare the value that a given mode frequency has at  $~t=0.2~$s and
at $~t=50~$s, we find a difference of about $~40\%~$ for the
f--mode,  $\sim~75\%~$ for the  p$_1$--mode and of $\sim 65\%$ for the  w$_1$--mode.
This significant difference can be attributed partly to the 
fact that at earlier times the PNS is less compact, and partly to the 
very reach thermodynamical structure that characterizes the first 
few seconds of the PNS life (see Figure \ref{FIG1}).
In this respect, the behavior of the  
the f--mode frequency is of particular interest, since it  does not scale 
with $\sqrt{M/R^3}$ as it does  for cold NSs: 
indeed, Tables \ref{table1} and \ref{table2} show that,
during the first second, the mass of the star 
remains approximately constant while the radius rapidly decreases,  
and consequently $\nu_f$ should  increase;
on the contrary, it  initially decreases, 
at about $~t\approx 0.5~$s approaches the value of the g--mode frequency, 
and only after about one second, it starts to increase linearly with $\sqrt{M/R^3}$.
At the same time, the g$_1$--mode frequency  increases
when the star is younger and hotter and
large thermal and composition gradients dominate in the interior,
it reaches a maximum value of about $~800-900~$ Hz within the first second of life, and
afterwards, as the star cools down and entropy gradients are smeared out,
it migrates towards lower values.
Even though there is a point of the evolution when $\nu_f$ and $\nu_{g_1}$
are very close (see Figure \ref{FIG3}),
the two modes  are always distinguishable 
because of the different number of nodes in their eigenfunctions.

From Tables \ref{table1} and \ref{table2}, we see that
after a few seconds from birth, the damping times of the f, p$_1$ and w$_1$--modes 
of the  PNS are quite close to those of its cold descendant. 
At very early times ($0.2 <t<2 $ s) the situation is different;
for instance, $\tau_f$  decreases rapidly from  $\sim 70~$ s to
$\sim 2-3~$ s, and it is interesting to compare  its values
with $\tau_{diss}$, which is about 10-20 seconds, 
as explained in Section 2.1.
For $ t \lappreq 0.5$ seconds,   $\tau_{diss} < \tau_f$, therefore during this time,
a small fraction of the amount of energy initially stored in the f--mode,
will be dissipated by neutrino processes.
However, for  $t \gappreq 0.5$ seconds, $\tau_f$ becomes smaller than $ \tau_{diss}$, 
and gravitational wave emission becomes the dominant dissipation mechanism.
In addition,  apart from the first few tens of milliseconds,
the p$_1$--mode damping time is always comparable to  $\tau_{diss}$, while
$~\tau_{w_1}~$ is basically unaffected by  the evolution and it is by far
the shortest one. Therefore, if some energy is initially stored into these modes, it will
be emitted in gravitational waves.

Furthermore, we find that during the first second,
the g$_1$--mode damping time is much shorter than that expected from
previous calculations for cold NSs. In particular, for both PNS models 
$~\tau_{g_1}~$ is comparable to $\tau_f,$ to  $\tau_{p_1}$
and also to the neutrino dissipation  timescale during the first half a second. 
Thus, the common belief, based on cold NSs dynamical behavior,
that the damping time of the f--mode is an
order of magnitude shorter than that of the first p--mode and many orders
of magnitude smaller than the g--modes damping times,
is no longer correct during the first second of life, 
which is likely to be the most important for the emission of GWs. 

For g--modes of order higher than one,  we find
damping times much larger than that of the g$_1$--mode; 
For instance, the g$_2$--mode has a damping time of the order of
$10^4~$s at $t=0.4~$s.  Thus, g--mode pulsations of order higher 
than 1  are  unlikely to be significant sources of gravitational 
radiation and will be neglected in our analysis.

\begin{figure}
\begin{center}
\centerline{\mbox{ 
\psfig{figure=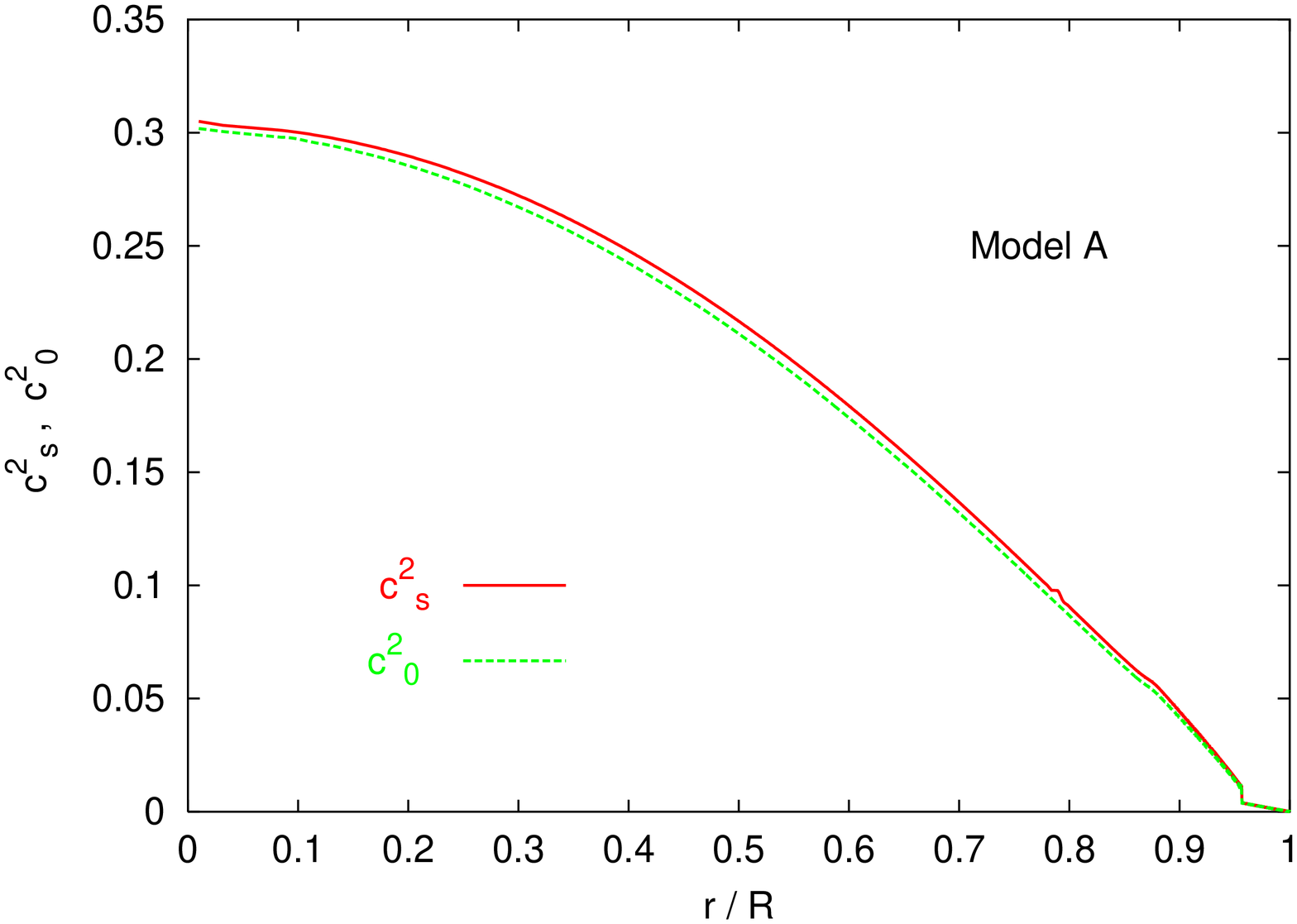,width=9.0cm,height=8.0cm,angle=-90}    
\psfig{figure=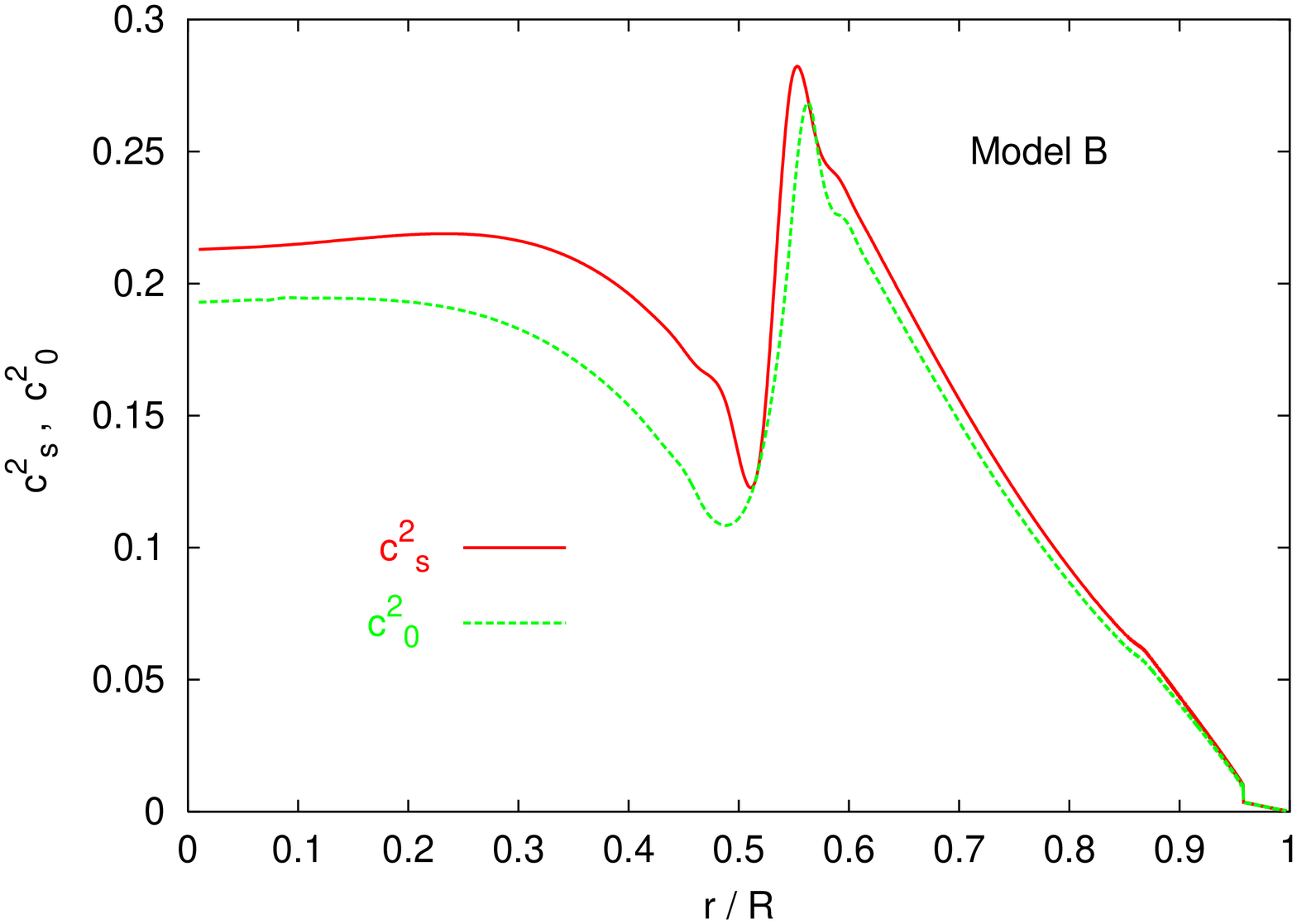,width=9.0cm,height=8.0cm,angle=-90}    
}}
\vspace{0.5cm}
\caption{
The square of the sound velocity $c_s$ and of the equilibrium velocity $c_0$ (see Section 3) 
for model A and B, as a function of the radial coordinate at $t=50~$ s. 
The difference between the  two stellar models
is due to the presence  of a  quark matter core in model B 
for $r < 0.54 R$. The larger difference between $c_s^2$ and $c_0^2$ in the 
quark core explains the higher value of $\nu_{g_1}$ for model B at the end of
the evolution.
}
\label{FIG4}
\end{center}
\end{figure}

From Tables \ref{table1} and \ref{table2} we see that, for
$ t\gappreq 10-20~$seconds, the g$_1$--mode frequency of both models increases.
As explained in Section 3, g--modes can be associated to several physical processes,
and during the early evolution of a PNS they arise  mainly because there are large
entropy and composition gradients. 
While during the deleptonization epoch entropy gradients dominate
the dynamical behavior of the star, after some time 
composition gradients become important, and this change
produces a modulation in the frequency of the g--modes.
In addition, at $t=50$ s we see that $\nu_{g_1}$ of model A and B are different. The reason
can be understood by looking at Figure \ref{FIG4}, where we plot 
$c_s^2$ and $c_0^2$ 
as a function of the radial distance at $t=50$ s.
For model A the difference between the two quantities is small
compared to model B, which develops a quark core and  an associated 
larger composition gradient.
As discussed in  Section 3, this explains why  the frequency of the g$_1$--mode 
for model B is larger than that of model A.

Finally,  it should be mentioned that
the frequency of the first w--mode ranges within $[1.4-1.8]$ kHz, which is  about
one order of magnitude smaller than the typical value of a cold NS, while the damping time
is similar, and very short ($\sim 10^{-4}$ s), much shorter than $\tau_{diss}$. 

\subsection{Gravitational wave  detectability}

The fact that during the early evolution of a PNS the frequencies of the quasi--normal
modes are much lower than those of the final NS  is of particular interest, since the
region where the interferometric antennas of first generation are more sensitive
does not extend much beyond 1 kHz.
This means that  waves emitted at the frequencies of the QNMs
of a cold NS will not be detectable (see Eqs. \ref{nilsfits}).

Unfortunately we do not know how the mechanical energy is distributed 
among the various modes,   but
we cannot exclude a priori that at very early times, besides the f--mode, the first g--mode 
or the first p--mode could give a contribution to the emitted wave, therefore 
we will consider each mode separately and we will proceed as follows. 
Having calculated how the frequencies and damping times evolve in time, we can model
the gravitational signal  emitted by the star oscillating in a given mode as
\begin{equation}
h(t) = {h_0}~{\mathrm e}^{-(t-t_0)/\tau(t)} \sin [~2\pi\nu(t) (t - t_0)~] \ ,
\label{h}
\end{equation}
where $~{h_0}~$ is the initial amplitude at the detector site, $~t_0~$ is the arrival time, 
$~\nu(t)~$ is the frequency of the oscillation, and $~\tau(t)~$ is 
the gravitational damping time of the oscillation. 
The signal--to--noise ratio (SNR) of a given detector 
is evaluated by using the standard formula for 
optimal matched filtering
\be
\label{SNR}
SNR= 2\left[
\int_0^\infty~d\nu~\frac{\vert \tilde h(\nu)\vert^2}{S_n(\nu)}
\right]^{1/2},
\ee
where $S_n(\nu)$ is the  noise power spectral density of the detector.
We shall use the noise curves of VIRGO I, LIGO I, GEO600 \cite{gris}, and that of 
EURO (http://www.astro.cf.ac.uk/geo/euro/),
an interferometric  detector with extremely high  sensitivity  in the kHz region
whose performances have been devised by a  joint panel of European experts.

We discuss the results of our calculations for model B, but similar results are 
obtained for model A. Since the emission of  gravitational waves is 
efficient only if the gravitational damping time $\tau (t)$ is comparable 
or shorter than the dissipative timescale $\tau_{diss}$,     
the gravitational signal produced by the various modes 
has been considered different from zero only if the condition 
$\tau (t) \lappreq \tau_{diss}$ is satisfied.  

In Table \ref{table3} we show the values of the initial amplitude $h_0$ 
that a given signal with the form (\ref{h})
should have to be detectable with a SNR=5.
In addition, we can estimate the amount of energy emitted by a source by
integrating over the frequency and the surface the expression of the energy flux
\be
\frac{dE_{\rm{GW}}}{dS d\nu} = \frac{\pi}{2}\,\nu^2\,|\,\tilde h(\nu )\,|^2 ,
\ee
In Table \ref{table3} we also give the energy $E_{GW}$ that should be emitted 
by a galactic source oscillating in a given mode  to reach SNR=5.
Notice that, since the energy scales as $D^2$, the energy required to detect 
the same signal emitted by a source in  the Virgo cluster (15 Mpc) with SNR=5
is simply given by $2.25 \times 10^{6} E_{GW}$.

From Table \ref{table3}, we see that a signal detectable by the first generation of
ground--based interferometers with SNR=5 must have
an initial amplitude larger than $10^{-22}$. For instance, $h_0=2.3\cdot 10^{-22}$ 
for VIRGO I, if emitted by an f--mode oscillation.
This amplitude corresponds to an emission of energy of 
$3.5 \times 10^{-8}~M_\odot c^2$, in the event of   a galactic Supernova.
This is consistent with recent studies of the
axisymmetric collapse of the core of a massive star,  
where  the energy emitted in gravitational waves, averaged over all the considered models, 
is estimated to range within $[3.6\cdot 10^{-8}-8.2\cdot 10^{-8}]~M_\odot c^2$ \cite{harry1}.
In the above reference, the corresponding average amplitude for a source at $10$~kpc
is about $1.7\times 10^{-20}$, and the average spectrum of gravitational emission  
is peaked at 930 Hz.

The advanced interferometer EURO would be able to detect signals with 
initial amplitude $h_0 \gappreq 8\cdot 10^{-25}$ with
SNR=5 (see Table \ref{table3}). 
For sources located at the distance of the Virgo cluster 
this corresponds to an emission energy of the order of $10^{-8}~M_\odot c^2$, 
with a multiplicative factor depending on the particular fluid mode. 
Thus, a detector like EURO would increase the detection rate for 
Supernova events from a few events per century to a few events per month.

The role of the spacetime w$_1$--mode deserves a separate discussion.
It is interesting to note that for the w--modes the amount of energy required for 
the GW to be detected is similar to that of fluid modes (see
Table \ref{table3}) despite the wave amplitude has to be larger. 
This is essentially due to the smaller
gravitational damping time, $\tau_{w_1}\sim 10^{-4}$ s. 
However, these modes are very weakly coupled to the fluid, and it remains
unclear whether this small amount of energy can be effectively conveyed
into spacetime oscillations.

\begin{figure}
\begin{center}
\psfig{figure=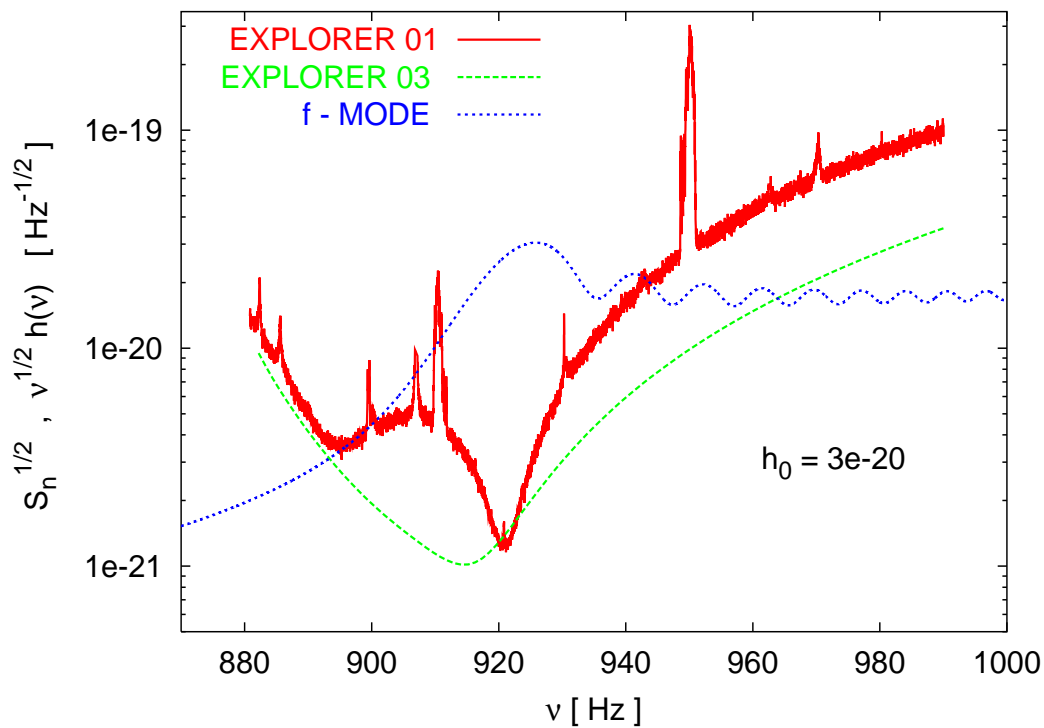,width=14.0cm,height=10.0cm,angle=-90}
\vspace{0.5cm}
\caption{
The  strain amplitude of the gravitational signal produced by  a proto--neutron star
oscillating in the f--mode is plotted as a function of the
frequency. The signal is assumed to have an initial  
amplitude of  $h_0 =  3 \times 10^{-20}$ (see Section 4).
In the same figure we also  plot the experimental noise strain amplitude  
of the resonant detector EXPLORER obtained in December 2001, and
the strain noise curve that the EXPLORER team expect to obtain in 2003.
}
\label{FIG5}
\end{center}
\end{figure}

We have also analyzed the response of a resonant mass detector to 
the gravitational signals computed in this paper.
In  Figure \ref{FIG5} we plot the experimental noise strain amplitude 
of EXPLORER obtained in December 2001 (http://www.roma1.infn.it/rog/explorer/),
the strain noise curve that the EXPLORER team expects to obtain in 2003 reducing the 
noise and increasing the quality factor \cite{pia},
and the strain amplitude produced by the f-- and  g$_1$--mode, corresponding to 
a SNR$=5$.
From the figure we see that only the f--mode signal falls in the region 
of best sensitivity. 
By using the 2003 noise curve, we find that it would be 
detected with SNR$=5$ if the amplitude on Earth
is $h_0 \simeq 3\cdot 10^{-20}$, i.e. $E_{GW}\simeq 6\cdot 10^{-4}~M_\odot c^2$ for a galactic source.
However,  this is only an indication; indeed, rotation or different details of the 
dynamical evolution of the star may shift the two contributions in a way that 
either the g--modes, or both modes  may fall inside the sensitivity band.
In addition, better sensitivities will be reached in the next future
by the bars further refining the experimental setting.
With the expected improvements (M. Bassan, private communication)
we find that $E_{GW}$ can be lowered by at least an order of magnitude.

\section{ Concluding Remarks}

In this paper we show that  the frequency 
and  the damping time of the QNMs  of a newborn proto--neutron star
change significantly as the star cools down and deleptonizes.
In particular, at early times, when most of the
gravitational energy is  emitted, the typical frequencies 
are considerably lower than those of cold NSs.
In order to understand whether the dissipative processes acting in a young PNS 
could damp the oscillations of the star more efficiently than gravitational waves,
we have compared the gravitational damping times to the 
timescales associated to neutrino viscosity,
diffusivity and thermal conductivity ($\tau_{diss} \approx 10 -20$ s).
We find that, during the first half a second, neutrino processes 
dominate with respect to the gravitational emission due to the f--mode.
However,  this time interval is very short compared to  $\tau_{diss}$,
and therefore no much of the energy initially stored into the mode will be 
dissipated  by neutrino diffusion. 
After half a second, the gravitational damping time of the f--mode 
becomes shorter than $\tau_{diss}$ 
and, consequently, the remaining mechanical energy can be 
freely released in gravitational waves. The p--mode damping time is always 
comparable to $\tau_{diss}$, while the gravitational emission 
from g--modes could be expected only during the first half a second 
of the evolution, when the g--mode damping time is comparable to 
$\tau_{diss}$.
The w--mode damping time is  always orders of magnitude smaller than
$\tau_{diss}$.

We have restricted our analysis to two models of PNS evolution, 
but we do not expect qualitative changes if different EOSs are used.
For example, by changing the symmetry energy of the equation of
state one can vary the  final composition gradient. 
For  changes in a physically reasonable range,
the g$_1$--mode frequency of the final, cold neutron star may vary
between 100-300 Hz  for the model without a phase transition, 
and higher values would be reached if a  
phase transition to quark matter occurs.
From our data (Tables \ref{table1} and \ref{table2}) 
we see that during the first second of evolution
$\nu_{g_1}$ varies within $\sim[600-800]$ Hz,  $\nu_f$ within
$\sim [900-1100]$ Hz, and 
$\nu_{g_1}$ reaches a maximum approximately when  $\nu_f$ has a minimum;
it is reasonable to expect that these features will not change significantly 
for different initial models.

As shown in Section 4.3, a gravitational signal, emitted at the 
frequency of the quasi--normal modes of a newborn proto--neutron 
star in the galaxy, could be detected 
with a signal to noise ratio of 5 by the first generation
interferometers if the energy stored in the modes
is greater than  $\sim 10^{-8}~M_\odot c^2$ 
(or by a resonant antenna if it is greater than $\sim 10^{-4}~M_\odot c^2$).
It is interesting to stress that the f--, p-- and w--modes have
frequencies much lower than those of the cold neutron star which forms at
the end of the evolution, in a region
where the detectors sensitivity is higher and the energy
threshold for detection is lower. 

Unfortunately, a galactic Supernova is a very rare event ($\approx 3$ per century).
However, there is another astrophysical scenario to which the results of our study could be applied.
Recent numerical simulations \cite{shibata1,shibata3} have shown 
that short-lived supra-massive neutron stars can also be formed
following the merger of two neutron stars with 
nearly equal masses and low compactness.
These highly non axisymmetric objects are supported by differential rotation
and would emit quasi-periodic gravitational waves
with typical frequencies of $\sim 2-3$ kHz. The results of these simulations show that
up to 0.01 $M_\odot$ could be radiated in gravitational
waves, as opposed to the 10$^{-8}$--10$^{-6}$ $M_\odot$ usually
obtained for stellar core collapse. In these works,
the coalescing stars are assumed to have a polytropic EOS 
and thermal effects are not included, thus the peak frequency corresponds to 
the excitation of the f-mode frequency of a cold neutron star. 
The fastly rotating PNS  born as a consequence of the coalescence of low compactness neutron
stars, should have thermodynamical properties similar to those 
of the PNSs we study in this paper. Therefore,
we think that the effects of thermal and composition gradients  
we have found should produce a similar shift in frequencies and damping times 
with respect to the cold star case, but with
the remarkable difference that these metastable PNSs would be much
stronger emitters of GWs.
The large amount of emitted  energy 
would make the ringing down of the PNS observable from the
Virgo cluster, increasing the event rate to a few per year.

\begin{center}
\bf{Acknowledgements}
\end{center}
We thank P. Astone for kindly providing the experimental curves of the noise of EXPLORER 
and for interesting discussions.

This work has been supported by the EU Programme `Improving the Human
Research Potential and the Socio-Economic Knowledge Base' (Research
Training Network Contract HPRN-CT-2000-00137) and the Spanish Ministerio
de Ciencia y Tecnologia grant AYA 2001-3490-C02-01.
JAP was supported by the Marie Curie Fellowship No. HPMF-CT-2001-01217.

\clearpage

\begin{table}
\begin{center}
\caption{
The frequencies (in Hz) and damping times (in s) of the first g--mode, 
the fundamental mode, the first p--mode, and the first w--mode are
given for different values of time (in s) for the PNS evolution, model
A. The damping time of the g$_1$--mode for $t>3~$ s is not shown because
it is so long that the oscillation will be damped by neutrinos 
rather than by GWs. As a reference, we indicate also the radius (in km) of
the PNS at each considered time step. The gravitational mass is $M = 1.58
M_\odot$ for our initial model at $t= 0.2$ s, and  $M = 1.46 M_\odot$ at
$t= 50$ s. The gravitational mass difference ($~\approx 0.12 M_\odot~$)
has been taken away by neutrinos during the PNS evolution.
}
\label{table1}
\vskip 5pt
\begin{tabular}{||p{0.001cm}*{10}{c|}|}
\hline
\hline
& $t$ & $R$  & $\nu_{g_1}$  & $\tau_{g_1}$  & $\nu_{f}$  & $\tau_{f}$ &
$\nu_{p_1}$  & $\tau_{p_1}$ & $\nu_{w_1}$ & $10^4
\times\tau_{w_1}$ \\
\hline
\hline
&$0.20$ & $34.2$ & $617.1$ & $21.8$ & $953.1$ & $60.2$ & $1221.5$ & $11.8$
& $1597.5$ & $2.00$ \\
\hline
& $0.25$ & $31.2$ & $677.0$ & $17.4$ & $948.9$ & $38.6$ & $1366.2$ &
$9.1$ & $1781.7$ & $1.92$ \\
\hline
& $0.30$ & $28.9$ & $727.0$ & $15.2$ & $943.8$ & $27.1$ & $1487.1$ & $7.7$
& $1943.3$ & $1.87$\\
\hline
&$0.40$ & $25.7$ & $796.4$ & $16.9$ & $941.6$ & $12.6$ & $1684.9$ & $6.6$
& $2213.5$ & $1.78$ \\
\hline
&$0.50$ & $23.7$ & $819.9$ & $35.8$ & $960.9$ & $7.1$ & $1833.5$ & $6.3$
& $2413.0$ & $1.73$ \\
\hline
&$0.60$ & $22.3$ & $815.4$ & $100.6$ & $994.4$ & $5.4$ & $1954.8$ & $6.2$ &
$2573.4$ & $1.69$ \\
\hline
& $0.70$ & $21.3$ & $799.5$ & $270.8$ & $1031.4$ & $4.7$ & $2057.6$ & $6.3$
& $2712.9$ & $1.66$ \\
\hline
& $0.75$ & $20.9$ & $790.3$ & $415.4$ & $1048.5$ & $4.4$ & $2100.4$ & $6.3$
& $2772.9$ & $1.65$ \\
\hline
& $1.0$ &$19.4$ & $744.1$ & $2.4\times 10^3$ & $1126.7$ & $3.6$ &
$2295.5$ & $6.8$ &$3033.6$ & $1.60$ \\
\hline
& $2.0$ & $15.7$ & $595.2$ & $2.5\cdot 10^5$ & $1358.5$ & $2.3$ & $3318.9$ &
$10.5$ & $3817.2$ & $1.50$ \\
\hline
&$3.0$ & $14.3$ & $486.2$ & $-$ & $1478.6$ & $1.9$ & $4087.7$ & $15.0$ &
$4238.2$ & $1.45$ \\
\hline
&$5.0$ & $13.6$ & $349.8$ & $-$ & $1565.0$ & $1.7$ & $4593.4$ & $19.5$ &
$4533.2$ & $1.42$ \\
\hline
&$10.0$ & $13.1$ & $159.1$ & $-$ & $1639.1$ & $1.6$ & $4983.8$ & $21.8$ &
$4775.8$ & $1.39$ \\
\hline
&$20.0$ & $12.9$ & $203.8$ & $-$ & $1678.2$ & $1.5$ & $5150.4$ & $20.4$ &
$4893.1$ & $1.37$ \\
\hline
&$30.0$ & $12.84$ & $229.3$ & $-$ & $1687.3$ & $1.5$ & $5192.3$ & $19.8$ &
$4918.2$ & $1.36$ \\
\hline
&$40.0$ & $12.81$ & $237.2$ & $-$ & $1689.7$ & $1.5$ & $5211.5$ & $20.0$ &
$4925.5$ & $1.36$ \\
\hline
&$50.0$ & $12.78$ & $242.2$ & $-$ & $1693.2$ & $1.5$ & $5228.5$ & $19.9$ &
$4936.4$ & $1.36$ \\
\hline
\hline
\end{tabular}
\end{center}
\end{table}

\clearpage      
\begin{table}
\begin{center}
\caption{
The same as in Table \ref{table1} but for model B. The main 
difference between the two models is that model A does not have a 
phase transition to quark matter in the inner core, 
while model B develops a quark 
core at about $t=20~$s.
The initial and final gravitational mass are the same as for model A.
}
\label{table2}
\vskip 5pt
\begin{tabular}{||p{0.001cm}*{10}{c|}|}
\hline
\hline
& $t$ & $R$  & $\nu_{g_1}$  & $\tau_{g_1}$  & $\nu_{f}$  & $\tau_{f}$ &
$\nu_{p_1}$  & $\tau_{p_1}$ & $\nu_{w_1}$ & $10^4
\times\tau_{w_1}$ \\
\hline
\hline
&$0.20$ & $35.1$ & $582.5$ & $24.5$ & $942.2$ & $69.2$ & $1145.1$ & $14.2$
& $1489.9$ & $2.06$ \\
\hline
& $0.25$ & $32.2$ & $636.0$ & $19.5$ & $939.0$ & $43.2$ & $1279.0$ &
$11.3$ & $1653.7$ & $1.98$ \\
\hline
& $0.30$ & $30.0$ & $682.0$ & $16.6$ & $934.2$ & $32.5$ & $1392.2$ & $9.7$
& $1796.3$ & $1.93$\\
\hline
&$0.40$ & $27.1$ & $750.0$ & $14.8$ & $926.2$ & $19.0$ & $1573.4$ & $8.3$
& $2028.0$ & $1.84$ \\
\hline
&$0.50$ & $25.1$ & $790.0$ & $18.6$ & $927.0$ & $10.9$ & $1718.0$ & $7.9$
& $2208.9$ & $1.79$ \\
\hline
&$0.60$ & $23.6$ & $805.3$ & $35.2$ & $941.4$ & $7.2$ & $1843.2$ & $7.8$ &
$2357.0$ & $1.75$ \\
\hline
& $0.70$ & $22.6$ & $803.4$ & $75.2$ & $962.6$ & $5.8$ & $1942.8$ & $7.9$
& $2471.5$ & $1.72$ \\
\hline
& $0.75$ & $22.2$ & $799.0$ & $108.0$ & $974.0$ & $5.4$ & $1985.0$ & $8.0$
& $2520.9$ & $1.71$ \\
\hline
& $1.0$ &$20.5$ & $767.0$ & $593.0$ & $1033.0$ & $4.4$ & $2181.0$ & $8.6$
&$2743.5$ & $1.66$ \\
\hline
& $2.0$ & $17.0$ & $640.0$ & $5\cdot 10^4$ & $1213.0$ & $2.9$ & $3036.0$ &
$13.5$ & $3369.3$ & $1.56$ \\
\hline
&$3.0$ & $15.3$ & $533.0$ & $-$ & $1316.0$ & $2.4$ & $3807.0$ & $21.0$ &
$3742.9$ & $1.52$ \\
\hline
&$5.0$ & $14.4$ & $374.0$ & $-$ & $1405.0$ & $2.1$ & $4437.0$ & $32.7$ &
$4050.1$ & $1.48$ \\
\hline
&$10.0$ & $13.9$ & $168.0$ & $-$ & $1475.5$ & $2.0$ & $4951.8$ & $43.0$ &
$4285.2$ & $1.44$ \\
\hline
&$20.0$ & $13.7$ & $232.6$ & $-$ & $1501.5$ & $1.9$ & $5144.8$ & $42.9$ &
$4364.6$ & $1.42$ \\
\hline
&$30.0$ & $13.5$ & $257.4$ & $-$ & $1547.1$ & $1.8$ & $5174.5$ & $28.2$ &
$4486.6$ & $1.40$ \\
\hline
&$40.0$ & $13.3$ & $476.2$ & $-$ & $1603.9$ & $1.7$ & $5231.2$ & $18.1$ &
$4640.6$ & $1.39$ \\
\hline
&$50.0$ & $13.2$ & $489.0$ & $-$ & $1606.9$ & $1.7$ & $5251.5$ & $17.9$ &
$4649.4$ & $1.38$ \\
\hline
\hline
\end{tabular}
\end{center}
\end{table}
\clearpage
\begin{table}
\begin{center}
\caption{The initial amplitude $h_0$  that a gravitational 
signal emitted by a PNS oscillating in the various modes should have, at the detector site,
to be detectable
by the VIRGO I, LIGO I, GEO600 and EURO interferometers, with a SNR$=5$.
We also show the corresponding energy $E_{GW}$ (in units of $M_\odot c^2$)
emitted in GWs for each mode, if the
source is located at a distance of $D=10$ kpc. Notice that the energy scales as $D^2$, and
the required energy to detect the same signal at the Virgo cluster (15 Mpc)
is  $2.25 \times 10^{6} E_{GW}$.}
\label{table3}
\vskip 5pt
\begin{tabular}{||p{0.001cm}*{6}{c|}|}
\hline
\hline
& QNM & & VIRGO I & LIGO I  & GEO600  & EURO  \\
\hline
\hline
& g$_1$--mode &  $h_0$ & $3.6\times 10^{-22}$ & $8.3\times 10^{-22}$ & $2.0\times 10^{-21}$ & $1.4\times 10^{-24}$ \\
&  &  $E_{GW}$ & $8.2\times 10^{-9}$ & $4.3\times 10^{-8}$ & $2.5\times 10^{-7}$ & $1.2\times 10^{-13}$ \\
\hline
& f--mode & $h_0$ & $2.3\times 10^{-22}$ & $6.2\times 10^{-22}$ & $1.6\times 10^{-21}$ & $8.2\times 10^{-25}$ \\
&  &  $E_{GW}$ & $3.5\times 10^{-8}$ & $2.7\times 10^{-7}$ & $1.7\times 10^{-6}$ & $4.6\times 10^{-13}$ \\
\hline
& p$_1$--mode & $h_0$ & $3.8\times 10^{-22}$ & $1.2\times 10^{-21}$ & $3.1\times 10^{-21}$ & $1.3\times 10^{-24}$ \\
&  &  $E_{GW}$ & $7.2\times 10^{-7}$ & $6.5\times 10^{-6}$ & $4.7\times 10^{-5}$ & $8.2\times 10^{-12}$ \\
\hline
& w$_1$--mode & $h_0$ & $2.8\times 10^{-20}$ & $4.6\times 10^{-20}$ & $9.3\times 10^{-20}$ & $1.1\times 10^{-22}$ \\
&  &  $E_{GW}$ & $5.9\times 10^{-8}$ & $1.6\times 10^{-7}$ & $6.1\times 10^{-7}$ & $9.5\times 10^{-13}$ \\
\hline
\hline
\end{tabular}
\end{center}
\end{table}



\clearpage

\clearpage
 
 \clearpage



\end{document}